\documentclass[preprint,aps,prd,showpacs,superscriptaddress,nofootinbib,tightenlines]{revtex4}
\usepackage{amsfonts}
\usepackage{multirow}
\usepackage{mathrsfs}
\usepackage{graphicx}
\usepackage{amsmath}
\usepackage{amssymb}
\usepackage{bm}
\usepackage{bbm}
\usepackage{color}
\usepackage{diagbox}

\def\pslash{p\!\!\!\slash}
\def\pbarslash{\bar{p}\!\!\!\slash}
\def\epslash{\epsilon\!\!\!\slash}
\def\Pslash{P\!\!\!\slash}

\def\OMIT#1{}

\newcommand{\nn}{\nonumber}

\newcommand{\beq}{\begin{equation}}
\newcommand{\eeq}{\end{equation}}
\newcommand{\bqa}{\begin{eqnarray}}
\newcommand{\eqa}{\end{eqnarray}}

\begin{document}
\title{\mbox{}\\[10pt]
${\cal O}(\alpha_sv^2)$ corrections to hadronic decay of vector quarkonia}
\author{Wen-Long Sang~\footnote{wlsang@swu.edu.cn}}
 \affiliation{School of Physical Science and Technology, Southwest University, Chongqing 400700, China\vspace{0.2cm}}

\author{Feng Feng\footnote{F.Feng@outlook.com}}
\affiliation{China University of Mining and Technology, Beijing 100083, China\vspace{0.2cm}}
\affiliation{Institute of High Energy Physics, Chinese Academy of
Sciences, Beijing 100049, China\vspace{0.2cm}}

\author{Yu Jia\footnote{jiay@ihep.ac.cn}}
\affiliation{Institute of High Energy Physics, Chinese Academy of
Sciences, Beijing 100049, China\vspace{0.2cm}}
\affiliation{School of Physics, University of Chinese Academy of Sciences,
Beijing 100049, China\vspace{0.2cm}}

\date{\today}

\begin{abstract}
Within the nonrelativistic QCD (NRQCD) factorization framework, we compute the ${\mathcal O}(\alpha_s v^2)$ corrections to the hadronic decay rate of vector
quarkonia, exemplified by $J/\psi$ and $\Upsilon$.
Setting both the renormalization and NRQCD factorization scales to be $m_Q$, we obtain
$\Gamma(J/\psi\to {\rm LH})=
0.0716\frac{\alpha_s^3}{m_c^2} \langle \mathcal{O}_1({}^3S_1)\rangle_{J/\psi} [1-1.19\alpha_s+(-5.32+3.03\alpha_s)\langle v^2\rangle_{J/\psi}]$ and
$\Gamma(\Upsilon\to {\rm LH})=
0.0716\frac{\alpha_s^3}{m_b^2}\langle\mathcal{O}_1({}^3S_1)\rangle_{\Upsilon}[1-1.56\alpha_s+(-5.32+4.61\alpha_s)\langle v^2\rangle_{\Upsilon}]$. We confirm the previous calculation of $\mathcal{O}(\alpha_s)$ corrections
on a diagram-by-diagram basis, with the accuracy significantly improved.
For $J/\psi$ hadronic decay, we find that the ${\mathcal O}(\alpha_sv^2)$ corrections are moderate
and positive, nevertheless unable to counterbalance the huge negative corrections.
On the other hand, the effect of ${\mathcal O}(\alpha_sv^2)$ corrections for $\Upsilon(nS)$
is sensitive to the $\mathcal{O}(v^2)$ NRQCD matrix elements.
With the appropriate choice of the NRQCD matrix elements, our theoretical predictions for
the decay rates may be consistent with the experimental data for $\Upsilon(1S,2S)\to {\rm LH}$.
As a byproduct, we also present the theoretical predictions for the branching ratio of $J/\psi(\Upsilon)\to 3\gamma$ accurate up to $\mathcal{O}(\alpha_s v^2)$.
\end{abstract}

\pacs{\it 12.38.Bx, 13.20.Gd, 13.40.Hq}

\maketitle

\section{Introduction}

The successful predictions of charmonium annihilation decay rates are among
the earliest triumph of perturbative QCD~\cite{Appelquist:1974zd}.
Among the quarkonia families, the spin-triplet $S$-wave sates with $J^{PC}=1^{--}$,
exemplified by  $J/\psi$ and $\Upsilon$, undoubtedly occupy the central stage,
whose various decay channels have been extensively studied both experimentally and theoretically.
Among a variety of decay channels of the vector quarkonia,
the inclusive hadronic decays are particularly interesting and important.
Obviously these are not only the most dominant decay channels, but an ideal place to test our understanding about
the interplay between perturbative and nonperturbative aspects of QCD.
Historically, $\Upsilon \to {\rm light\;hadrons}$ has been employed to calibrate
the strong coupling constant at the scale of bottom mass~\cite{Mackenzie:1981sf,Brambilla:2007cz}.

At lowest order, the inclusive hadronic decays of vector quarkonia proceed via $J/\psi(\Upsilon)\to 3g$, as demanded
by conservation of $C$ parity. This is very similar to the ortho-positronium (o-Ps) annihilation
decay into three photons~\cite{Caswell:1976nx}, so the corresponding expression can be directly transplanted
supplemented with the proper color factor.

Since quarkonia are essentially the bound states composed of slowly-moving
heavy quark and heavy antiquark,
the consensus is nowadays that quarkonium annihilation decay processes can be reliably
tackled in nonrelativistic QCD (NRQCD) factorization framework~\cite{Bodwin:1994jh},
which effectively organizes the theoretical predictions as double expansion in $\alpha_s$ and $v$.
The $\mathcal{O}(\alpha_s)$ perturbative corrections, which turn out to be negative,
was first computed by Mackenzie and Lepage in 1981~\cite{Mackenzie:1981sf}.
In sharp contrast to the significantly negative $\mathcal{O}(\alpha_s)$  corrections in $J/\psi\to 3\gamma$,
the $\mathcal{O}(\alpha_s)$ correction in $J/\psi(\Upsilon)$ hadronic decay appears to be moderate in magnitude.
On the other hand, the leading relativistic corrections to $J/\psi(\Upsilon)\to 3g,\,3\gamma$
were first calculated by Keung in 1982~\cite{Keung:1982jb}. With reasonable assumption of the
relativistic NRQCD matrix elements for $J/\psi$, the $\mathcal{O}(v^2)$ corrections appear to
be significantly negative for both $J/\psi(\Upsilon)$ inclusive hadronic decay channel and three-photon channel.
Bodwin {\it et al.} proceeded further to compute the $\mathcal{O}(v^4)$ corrections to the $J/\psi(\Upsilon)$ hadronic decay~\cite{Bodwin:2013zu}. It was observed that the relativistic corrections in the color-singlet channel exhibit decent convergence pattern,
but for consistency one should also include contributions from
various color-octet operator matrix elements at $\mathcal{O}(v^4)$.
Unfortunately, the actual values of the higher-order NRQCD color-octet matrix elements are rather poorly known.
The proliferation of many poorly constrained nonpertubative matrix elements
severely hinders the predictive power of NRQCD approach~\footnote{Although the NRQCD matrix elements
of the color-octet production operators have been fitted in various places~\cite{Butenschoen:2011yh,Chao:2012iv,Sun:2015pia},
there does not exist rigorous relation between the color-octet production matrix elements and
the decay matrix elements~\cite{Bodwin:1994jh}.}.

It is interesting to recall how a longstanding puzzle for  $J/\psi\to 3\gamma$ is resolved.
After incorporating both substantially negative $\mathcal{O}(\alpha_s)$ and $\mathcal{O}(v^2)$ corrections,
one simply ends up with a negative, hence unphysical, prediction to the partial width!
Fortunately, the dilemma is greatly reconciled after including the joint perturtative and relativistic correction, {\it e.g.},
the $\mathcal{O}(\alpha_sv^2)$ correction, which turns out to be positive and unexpectedly sizable~\cite{Feng:2012by}.
Incorporating this new piece of correction,
with the renormalization scale in the range $1.2\; {\rm GeV}< \mu< 1.4\; {\rm GeV}$,
one can reach satisfactory agreement between the state-of-the-art
NRQCD prediction~\cite{Feng:2012by} and the latest measurement at
{\tt BESIII}~\cite{Ablikim:2012zwa} for the rare decay channel $J/\psi\to 3\gamma$.

Inspired by the important role played by the $\mathcal{O}(\alpha_sv^2)$ correction to $J/\psi\to 3\gamma$,
the aim of this work is to investigate the $\mathcal{O}(\alpha_sv^2)$ correction to the  $J/\psi(\Upsilon)$ hadronic widths.
Obviously, the corresponding calculation is much more challenging than that for $J/\psi\to 3\gamma$.
We note that this joint radiative and relativistic correction is formally of the
comparable magnitude with the ${\cal O}(\alpha_s^2 v^0)$ and ${\cal O}(\alpha_s^0 v^4)$
corrections. Nevertheless, unlike these two types of corrections, which are either beyond current calculational
capability or lacking predictive power, we are equipped with sufficient technicality to tackle the $\mathcal{O}(\alpha_sv^2)$ correction,
and can also make some unambiguous and concrete predictions.
We hope our calculation can serve to further test the validity of NRQCD factorization approach in these basic vector quarkonia decay channels.

The remainder of the paper is organized as follows.
In Section~\ref{sec2-notations},
we recapitulate the NRQCD factorization formula for the hadronic decay of vector quarkonia and
set up the notations.
In Sec.~\ref{Strategy:calculation}, we sketch the calculational strategy for the intended NRQCD SDCs.
In Sec.~\ref{SDC:numerical}
we present the main numerical results. In particular, a detailed comparison with the existing
$\mathcal{O}(\alpha_s)$ corrections in a diagram-by-diagram basis has  also been given.
We devote Section~\ref{Section:Phenomenological:Analysis} to a comprehensive phenomenological analysis.
Finally we summarize in Section~\ref{summary}.
We also dedicate Appendix~\ref{phase:space:integration} to describe
the technicality in treating multi-body phase space integration in dimensional regularization.

\section{NRQCD factorization formula for hadronic width of vector quarkonia~\label{sec2-notations}}

In line with the NRQCD factorization~\cite{Bodwin:1994jh},
the annihilation hadronic decay rate of a vector quarkonium $V$ ($V$ may refer to $\Upsilon$ or $J/\psi$)
can be expressed as
\bqa
\label{NRQCD:factorization:formula}
\Gamma(V\to {\rm LH})&=&
F_1({}^3S_1)\langle \mathcal{O}_1({}^3S_1)\rangle_V
+\frac{G_1({}^3S_1)}{m_Q^2}\langle \mathcal{P}_1({}^3S_1)\rangle_V+{\mathcal{O}(v^4\Gamma)},
\eqa
where LH denotes the abbreviation for light hadrons, and
$m_Q$ is the heavy quark mass with quark species $Q=c,b$.
The four-fermion NRQCD operators in \eqref{NRQCD:factorization:formula} are defined by
\begin{subequations}
 \bqa
 \label{LDME-1}
\mathcal{O}_1({}^3S_1)&=&\psi^\dagger\bm{\sigma}\chi\cdot\chi^\dagger\bm{\sigma}\psi,
\\
\mathcal{P}_1({}^3S_1)&=&\frac{1}{2}\bigg[
\psi^\dagger\bm{\sigma}\chi\cdot
\chi^\dagger\big(-\tfrac{i}{2}\overleftrightarrow{\bm{D}}\big)^2\bm{\sigma}\psi
+
\psi^\dagger
\big(-\tfrac{i}{2}\overleftrightarrow{\bm{D}}\big)^2\bm{\sigma}\chi\cdot
\chi^\dagger\bm{\sigma}\psi
\bigg],
\eqa
\end{subequations}
whose matrix elements are expected to obey the velocity counting rule and
scale as $v^3$ and $v^5$, respectively.
Hence, the second term constitutes the ${\cal O}(v^2)$ relativistic correction
to the first term in \eqref{NRQCD:factorization:formula}.

The coefficients $F$ and $G$ in \eqref{NRQCD:factorization:formula}
are referred to as the NRQCD short-distance coefficients (SDCs), which
capture the relativistic ($k\sim 1/m_Q$) short-distance effects of QCD.
Owing to asymptotic freedom, they can be reliably computed in perturbation theory:
\begin{subequations}
\bqa
F_1({}^3S_1)=F_1^{(0)}({}^3S_1)+\frac{\alpha_s}{\pi}F_1^{(1)}({}^3S_1)+\cdots,\\
G_1({}^3S_1)=G_1^{(0)}({}^3S_1)+\frac{\alpha_s}{\pi}G_1^{(1)}({}^3S_1)+\cdots .
\eqa
\end{subequations}

The various SDCs can be deduced via the standard {\it
perturbative matching} technique. The key idea is because the SDCs like $F$, $G$ are insensitive to
the long-distance physics, we can replace the physical vector quarkonium state $V$
in \eqref{NRQCD:factorization:formula} with a fictitious quarkonium state
carrying the same quantum number ${}^3S_1$, yet composed of a free heavy quark-antiquark pair.
Concretely speaking, our matching equation reads
\bqa
\label{perturbative:matching:formula}
\Gamma(Q\overline{Q}(^3S_1)\to {\rm LH})&=&
F_1({}^3S_1)\langle \mathcal{O}_1({}^3S_1)\rangle_{Q\overline{Q}(^3S_1)}
+\frac{G_1({}^3S_1)}{m_Q^2}\langle \mathcal{P}_1({}^3S_1)\rangle_{Q\overline{Q}(^3S_1)}+{\mathcal{O}(v^4\Gamma)}.
\eqa
We can calculate both sides of \eqref{perturbative:matching:formula} using perturbative QCD and NRQCD,
then iteratively solve for a variety of SDCs to a prescribed order in $\alpha_s$.
Our main task in this work is to determine the unknown coefficient $G_1({}^3S_1)$.

\section{Technical strategy of calculating NRQCD SDCs\label{Strategy:calculation}}

\subsection{A shortcut to deduce the SDC at ${\cal O}(v^2)$\label{shortcut:deduce:SDC}}

In this section we sketch some important technicalities encountered in the perturbative matching calculation.
For the fictitious vector quarkonium appearing in \eqref{perturbative:matching:formula},
we assign the momenta carried by the heavy quark $Q$ and antiquark $\overline{Q}$ to be
\bqa
\label{kinematics-momenta}
p&=&\frac{P}{2}+q, \qquad
\bar{p}=\frac{P}{2}-q,
\eqa
where $P$ and $q$ denote the total momentum and half of the relative momentum of the $Q\overline{Q}$ pair, respectively.
The on-shell condition indicates
 \bqa
\label{kinematics-on-shell}
&&p^2=\bar{p}^2=m_Q^2,\qquad P\cdot q=0,\qquad P =4E^2,
\eqa
with $E=\sqrt{m_Q^2- q^2}>m_Q$.
Moreover, we assign the momentum of each massless parton in the decay products of $V$ to be
$k_i$ ($i=1,2,3$ for three particles in the final state, and $i=1,2,3,4$ for four-body decay),
which is subject to the on-shell condition $k_i^2=0$.

To facilitate the perturbative QCD calculation on the
left-hand side of \eqref{perturbative:matching:formula},
it is convenient to employ the covariant projector to enforce the $Q\overline{Q}$ pair
to be in the spin-triplet and color-singlet state.
The nonrelativistically-normalized spin-triplet/color-singlet
projector reads~\citep{Bodwin:2013zu}
 \bqa
\label{spin-projector}
\Pi_1=\frac{-1}{2\sqrt{2}E(E+m_Q)}(\pslash+m_Q)\frac{\Pslash+2E}{4E}\epslash(\pbarslash-m_Q)\otimes
{\textbf{1}_C\over \sqrt{N_c}},
\eqa
where $\epsilon^\mu$ represents the spin-1 polarization vector.

With the aid of \eqref{spin-projector}, the amplitude for the spin-triplet/color-singlet $Q\overline{Q}$ pair
to annihilating into massless partons can be obtained through
 \beq
\label{Spin:triplet:amplitude}
\mathcal{A}= {\rm Tr}[\widetilde{A}\Pi_1],
\eeq
where $\widetilde{A}$ denotes the quark amplitude for $Q\overline{Q} \to {\rm LH}$ with the
external quark spinors amputated, and it is understood
that the trace acts on both spinor and color indices.

We proceed to project out the $S$-wave piece from \eqref{Spin:triplet:amplitude}. To our purpose, we need expand
the amplitude  $\mathcal{A}$ through the second order in $q$.
The desired amplitude for $Q\overline{Q}(^3S_1)\to {\rm LH}$ can be expanded into
\bqa
\label{3S1:QQbar:Amplitude}
\mathcal{A}_S=\mathcal{A}_{S0}+\mathcal{A}_{S2}+\mathcal{O}(v^4),
\eqa
with
\begin{subequations}
\label{S-wave:Ampl:expansion}
\bqa
\mathcal{A}_{S0} &=& \mathcal{A}\big|_{{\bm q}\to 0},
\\
\mathcal{A}_{S2} &=& {{\bm q}^2\over 2(d-1)} \mathcal{I}^{\mu\nu}
{\partial^2 {\mathcal A} \over \partial q^\mu \partial q^\nu} \bigg|_{{\bm q}\to 0}.
\eqa
\end{subequations}
Here ${\bm q}^2= -q^2$ is defined in the rest frame of $Q\overline{Q}$ pair, and the symmetric
tensor is given by
 \bqa
\label{amps-s-wave-expand-explicit}
\mathcal{I}^{\mu\nu}=-g^{\mu\nu}+\frac{P^\mu P^\nu}{4E^2}.
\eqa
The above $\mathcal{A}_{S0}$ and $\mathcal{A}_{S2}$ represent the $\mathcal{O}(v^0)$ and $\mathcal{O}(v^2)$ amplitudes respectively.

Squaring the $S$-wave amplitude in \eqref{3S1:QQbar:Amplitude} and integrating over multi-body phase space,
truncating through order-${\bm q}^2$, we then accomplish the calculation of the
quark-level decay rate on left side of \eqref{perturbative:matching:formula}.
Nevertheless, a technical subtlety may be worth mentioning. Firstly, some portion of
relativistic correction is hidden in the phase space integral, since the invariant mass of $Q\overline{Q}$ is $4E^2$ instead of $4m_Q^2$.
Secondly, the squared $S$-wave amplitude explicitly involves factors $P\cdot k_i$, which also depend on
the variable $E$ other than $m_Q$. As pointed out in Refs.~\cite{Jia:2009np,Feng:2012by}, in order to extract the relativistic
correction, it may be more convenient to expand the quark amplitude \eqref{Spin:triplet:amplitude}
in power series of $\tfrac{{\bf q}^2}{E^2}$.
This strategy guarantees that the relativistic effect is automatically taken into account in the phase space integral.
Finally, we return to more conventional expressions by further expanding $E=\sqrt{m_Q^2+{\bm q}^2}$ in power of
$v^2\equiv\tfrac{{\bm q}^2}{m_Q^2}$.

Another point in calculating the virtual correction is also worth mentioning.
Prior to conducting loop integration, we have already expanded the integrand of the quark amplitude in power series of ${\bm q}^2$.
This operation amounts to directly extracting the hard matching coefficient in the context of strategy of region~\cite{Beneke:1997zp}. Therefore,
we are no longer concerned with Coulomb singularity and other soft contributions ubiquitously arising in virtual correction calculation.
As a consequence, calculating the radiative correction to the NRQCD side of \eqref{perturbative:matching:formula} is
no longer required.
It suffices to know the following perturbative NRQCD matrix elements at leading order (LO) in $\alpha_s$:
\begin{subequations}
\label{NRQCD:pert:Matrix:elements}
\bqa
\langle \mathcal{O}_1({}^3S_1)\rangle_{Q\overline{Q}(^3S_1)}&\equiv&
\langle Q(p)\overline{Q}(\bar{p})(^3S_1) \vert \mathcal{O}_1(^3S_1) \vert Q(p)\overline{Q}(\bar{p})(^3S_1) \rangle =2N_c,
\label{NRQCD:pert:Matrix:elements:v0}
\\
\langle \mathcal{P}_1({}^3S_1)\rangle_{Q\overline{Q}(^3S_1)}&\equiv &
\langle Q(p)\overline{Q}(\bar{p})(^3S_1) \vert \mathcal{P}_1(^3S_1) \vert Q(p)\overline{Q}(\bar{p})(^3S_1) \rangle =2N_c {\bm q}^2.
\label{NRQCD:pert:Matrix:elements:v2}
\eqa
\end{subequations}
In a sense, rather than literally follow perturbative matching calculation,
we take an efficient shortcut to obtain the NRQCD SDCs.

\subsection{Strategy of calculating virtual and real corrections}

To coherently unify the treatment of the virtual and real corrections to $Q\overline{Q}({}^3S_1)$ inclusive decay,
we utilize the optical theorem to organize the calculation for ${\cal O}(\alpha_s)$ correction.
Concretely, we follow \cite{Mackenzie:1981sf} to consider the imaginary part of the $Q\overline{Q}({}^3S_1)$ forward-scattering amplitude,
and place all possible cuts in accordance with Cutkosky's rule.

\begin{figure}[htbp]
 \centering
 \includegraphics[width=1\textwidth]{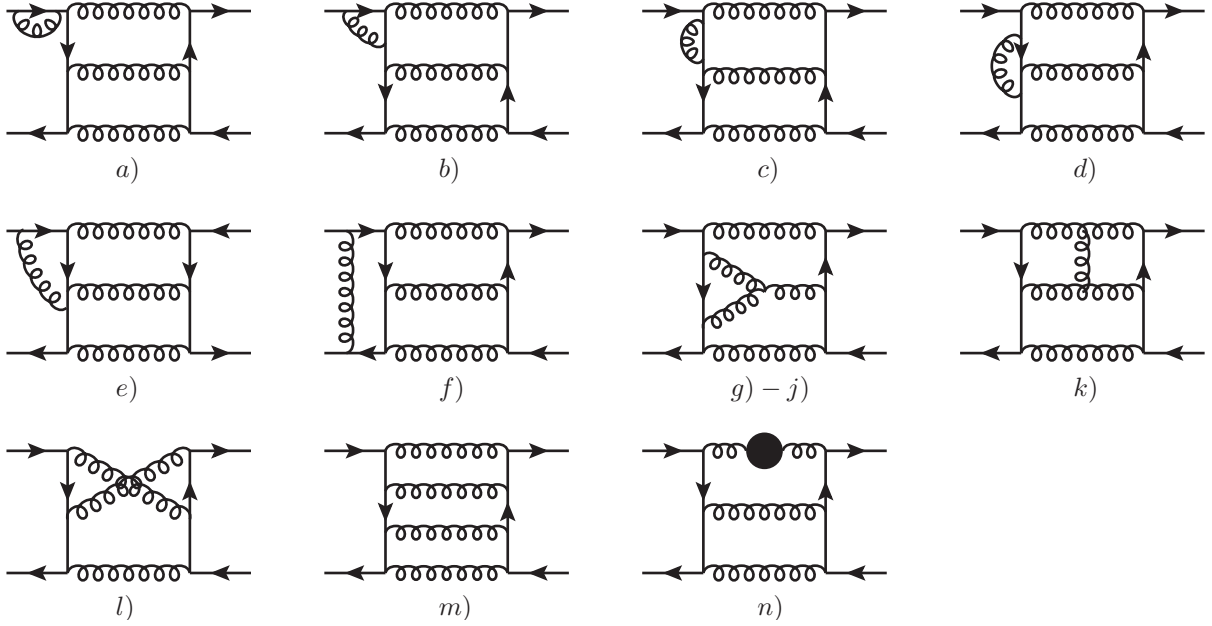}
 \caption{The relevant Feynman diagrams for the forward-scattering amplitude of
 $Q\overline{Q}({}^3S_1^{(1)})\to Q\overline{Q}({}^3S_1^{(1)})$ at ${\cal O}(\alpha_s^4)$. It is tacitly
 understood that a Custkosky cut should be placed on all possible intermediate state for each diagram, since
 we are only interested in the imaginary part of the amplitude.
 We follow the same classification convention as Ref.~\cite{Mackenzie:1981sf}.
 The solid blob in class $n)$ signifies the one-loop vacuum polarization diagrams that are composed of gluon and
 massless quarks.
 \label{feynman-fig-jpsi}}
 \end{figure}

Throughout the work, we choose Feynman gauge to compute the quark amplitude, and adopt the
dimensional regularization to regularize both UV and IR divergences.
We use the package {\tt FeynArts}~\cite{Hahn:2000kx} to generate Feynman diagrams and the corresponding
$Q\overline{Q}$ amplitudes.
As illustrated in Fig.~\ref{feynman-fig-jpsi},
all Feynman diagrams are categorized into a dozen of groups with distinct topology.
The Cutkosky's cut is implicitly assumed and acts on all possible places.
For each side of the cut, we apply the spin/color projector \eqref{spin-projector} together with
\eqref{Spin:triplet:amplitude}, \eqref{3S1:QQbar:Amplitude} and
\eqref{S-wave:Ampl:expansion}  to
project out the amplitude for $Q\overline{Q}({}^3S_1)$ transitioning into massless partons through
$\mathcal{O}(v^2)$. The packages {\tt FeynCalc}~\cite{Mertig:1990an} and {\tt Form}~\cite{Kuipers:2012rf}/{\tt FormLink}~\cite{Feng:2012tk}
are utilized for tensor contraction to obtain the squared $S$-wave amplitudes.

We utilize the packages {\tt Apart}~\cite{Feng:2012iq} to conduct partial fraction and {\tt FIRE}~\cite{Smirnov:2014hma}
for the corresponding IBP reduction. We end up with obtaining 17 MIs for the real corrections
and 58 MIs for the virtual corrections. For all the MIs, we use the packages
{\tt FIESTA}~\cite{Smirnov:2013eza}/SecDec~\cite{Carter:2010hi} to perform sector decomposition (SD)~\footnote{
To improve efficiency, we have interchanged the order of the operations for contour deformation~\cite{Borowka:2012qfa}, SD
and series expansion in the {\tt FIESTA}. For more technical details, we refer the interested readers to Refs.~\cite{Feng:2019zmt,Sang:2020fql}.}.
For each decomposed sector, we first use {\tt CubPack}~\cite{CubPack}/{\tt Cuba}~\cite{Hahn:2004fe} to
conduct the first-round rough numerical integration.
For those integrals with large estimated errors, a parallelized integrator {\tt HCubature}~\cite{HCubature} is utilized to repeat the numerical integration until the prescribed accuracy is achieved.

Both UV and IR divergences could arise in the NLO correction calculation.
The UV divergence is affiliated only to the virtual correction calculation.
We implement the on-shell renormalization for the heavy quark wave function and mass,
and renormalize the strong coupling constant under $\overline{\rm MS}$ prescription.
The virtual correction then becomes UV finite after standard renormalization procedure.
Nevertheless, both virtual and real corrections are still plagued with IR divergences.
At $\mathcal{O}(\alpha_sv^0)$, upon summing both contributions,
the IR divergences are exactly cancelled~\cite{Mackenzie:1981sf}.
As we shall see in Section~\ref{SDC:numerical}, a peculiar pattern actually emerges at $\mathcal{O}(\alpha_sv^2)$.
Upon summing virtual and real corrections, the IR singularities do not exactly cancel away, but leave
a logarithmic IR divergence as the remnant. This single IR pole should in fact be factored into the NRQCD matrix element,
so that one finally arrives at IR-finite yet scale-dependent SDC at ${\cal O}(\alpha_s v^2)$.

Some remarks are in order for treating the multi-body phase space integration.
Rather than directly integrating the squared amplitudes over the three(four)-body phase space,
which appears to be a daunting task, we employ some modern technique, {\it e.g.}, the reverse unitarity method~\cite{Anastasiou:2002yz,Gehrmann-DeRidder:2003pne} to simplify the calculation.
The trick is to convert a phase-space integral into a loop integral,
which is facilitated by the following identity involving the $i$-th cut propagator:
\beq
\int\!\! \frac{d^D k_i}{(2\pi)^{D}} 2 \pi\,i\,\delta(k_i^2)\theta(k_i^0) =
\int\!\! \frac{d^D k_i}{(2\pi)^{D}} \left(\frac{1}{k_i^2 + i\varepsilon} - \frac{1}{k_i^2 - i\varepsilon} \right).
\label{cut:trick}
\eeq
Since the differentiation operation is insensitive to the $\pm i\varepsilon$ in the propagator,
one can fruitfully apply the integration-by-parts (IBP) identities, which are
widely used in reducing the multi-loop integrals into a set of simpler Master Integrals (MIs),
also to reduce the various phase-space integrals into a set of simple MIs.
We then perform the corresponding multi-body phase space integration  numerically  with
these much simpler MIs.
In Appendix~\ref{phase:space:integration}, we provide some details on how to parameterize the
multi-body phase space integration in dimensional regularization.

\section{Numerical results for NRQCD SDCs
\label{SDC:numerical}}

In this section, we collect the known results of various SDCs in the NRQCD factorization formula
for $J/\psi(\Upsilon)$ hadronic decay, and report the major new result of this work, the ${\cal O}(\alpha_s v^2)$ correction
to SDC.

\subsection{Leading-order NRQCD SDC}

At LO in $\alpha_s$, the inclusive hadronic decay of the $Q\overline{Q}(^3S_1)$ pair
is characterized by its annihilation into three gluons.
It is a straightforward exercise to deduce the LO decay rate in $d=4-2\epsilon$ spacetime dimensions:
\bqa
\Gamma^{(0,0)}(Q\overline{Q}(^3S_1)\to 3g)&=&\frac{1}{3}\frac{8(4E^2)^{-1-2\epsilon}\alpha_s^3(4\pi)^{2\epsilon}}{\Gamma(2-2\epsilon)}
\bigg[\frac{20(\pi^2-9)}{27}\nn\\
&+&\frac{5}{36}\bigg(28+224\zeta(3)-29 \pi^2\bigg)\epsilon\bigg].
 \label{decay-pert-0-0}
\eqa
We have introduced the symbol $\Gamma^{(i,j)}$ to denote the
perturbative decay rate computed at a specific order in $\alpha_s$ and $v$ expansion,
{\it e.g.}, with the superscript $(i,j)$ signifying the joint ${\cal O}(\alpha_s^i v^{2j})$ correction.
For future usage, we have deliberately kept the ${\cal O}(\epsilon)$ piece in (\ref{decay-pert-0-0}).

Substituting (\ref{decay-pert-0-0}) and \eqref{NRQCD:pert:Matrix:elements:v0} into \eqref{perturbative:matching:formula},
and setting $\epsilon\to 0$, we then solve the SDC at ${\cal O}(\alpha_s^0 v^{0})$:
 \bqa
F_1^{(0)}({}^3S_1)&=&\frac{20\alpha_s^3}{243m_Q^2}
\left(\pi^2-9\right).
 \label{sdcs-0-0}
\eqa
In deriving this, we have replaced $E$ by $m_Q$, which is valid at LO in $v^2$. As expected,
we have just reproduced the classical result~\cite{Mackenzie:1981sf,Bodwin:1994jh,Bodwin:2013zu}.

\subsection{SDC at $\mathcal{O}(\alpha_s^0v^2)$}

Following the shortcut to extract the relativistic correction, as outlined in Section~\ref{shortcut:deduce:SDC},
we find the $\mathcal{O}(\alpha_s^0v^2)$ correction to the perturbative decay rate to be
 \bqa
\Gamma^{(0,1)}(Q\overline{Q}(^3S_1)\to 3g)&=&\Gamma^{(0,0)}(Q\overline{Q}(^3S_1)\to 3g) {{\bm q}^2\over E^2} \frac{24-7\pi^2}{12(\pi^2-9)}.
\label{decay-pert-0-1}
\eqa

To determine the SDC $G_1$, we need also expand $E$ in (\ref{decay-pert-0-1}) to first order in $\tfrac{{\bm q}^2}{m_Q^2}$.
Substituting (\ref{decay-pert-0-1}) and \eqref{NRQCD:pert:Matrix:elements:v2} into \eqref{perturbative:matching:formula},
with the aid of the known LO SDC $F_1^{(0)}$ in \eqref{sdcs-0-0}, we then deduce the SDC at ${\cal O}(\alpha_s^0 v^{2})$:
 \bqa
G_1^{(0)}({}^3S_1)&=&F_1^{(0)}({}^3S_1)\bigg[\frac{24-7\pi^2}{12(\pi^2-9)}-1\bigg]=F_1^{(0)}({}^3S_1)\frac{132-19\pi^2}{12(\pi^2-9)},
 \label{sdcs-0-1}
\eqa
which agrees with Refs.~\cite{Keung:1982jb,Bodwin:1994jh,Bodwin:2013zu}.

\subsection{SDC at $\mathcal{O}(\alpha_sv^0)$}

The QCD radiative correction to the vector quarkonia hadronic decay, yet at LO in velocity expansion,
was first computed by Mackenzie and Lepage nearly forty years ago~\cite{Mackenzie:1981sf}.
Dividing all the cut diagrams into several groups of distinct topology, the authors of ~\cite{Mackenzie:1981sf}
tabulate the numerical result of each individual class of diagrams, some of which have rather limited accuracy.
In order to make a close comparison with their result, we also adopt the same classification convention of Feynman diagrams
as \cite{Mackenzie:1981sf}.

\begin{table}
\caption{$\mathcal{O}(\alpha_sv^0)$ contribution to the $\Gamma^{(1,0)}/(\alpha_s\Gamma^{(0,0)}/\pi)$ from each class of cut diagrams, with our results
juxtaposed with the counterparts given in \cite{Mackenzie:1981sf}.
We have shifted the 't Hooft unit mass according to $\mu^2 \to \mu^2  e^{\gamma_E}/(4\pi)$,
which is equivalent to renormalizing $\alpha_s$ in the $\overline{\rm MS}$ scheme.
For simplicity, we temporarily set the renormalization scale $\mu_R=2E$. }
\label{table-as-v0}
\begin{ruledtabular}
\begin{tabular}{c|c|c|c|c}
Cut diagrams & Virtual corr. & Real corr. & Real+Virtual Corr. & Result from \cite{Mackenzie:1981sf}
\\
\hline
$a$) & $-\frac{2}{\epsilon}-5.439$ & --- & $-\frac{2}{\epsilon}-5.439$ & $-\frac{2}{\epsilon}-5.439$
\\
\hline
$b$) & $-\frac{1}{6\epsilon}-0.726$ & --- & $-\frac{1}{6\epsilon}-0.726$ & $-\frac{1}{6\epsilon}-(0.726\pm 0.002)$
\\
\hline
$c$) & $-\frac{4}{3\epsilon}-0.802$ & --- & $-\frac{4}{3\epsilon}-0.802$ & $-\frac{4}{3\epsilon}-(0.802\pm 0.002)$
\\
\hline
$d$) & $-\frac{1}{12\epsilon}-0.142$ & --- & $-\frac{1}{12\epsilon}-0.142$ & $-\frac{1}{12\epsilon}-(0.143\pm 0.001)$
\\
\hline
$e$) & $0.596$ & --- & $0.596$ & $0.594\pm0.002$
\\
\hline
$f$) & $\frac{4}{3\epsilon}-8.562$ & --- & $\frac{4}{3\epsilon}-8.562$ & $\frac{4}{3\epsilon}-(8.576\pm0.022)$
\\
\hline
$g$)-$j$) & $\frac{4.216}{\epsilon}-17.682$ & $\frac{2.534}{\epsilon}+21.318$ & $\frac{6.75}{\epsilon}+3.636$ & $\frac{6.75}{\epsilon}+3.54(30)$
\\
\hline
$k$) & $-\frac{4.5}{\epsilon^2}-\frac{8.767}{\epsilon}+13.921$ & $\frac{4.5}{\epsilon^2}+\frac{8.767}{\epsilon}-3.461$ & {$10.460(7)$} & $10.59\pm0.26$
\\
\hline
$l$) & -3.025 & --- & -3.025 & $-3.02\pm0.04$
\\
\hline
$m$) & --- & -0.181 & -0.181 &  $-0.19\pm0.04$
\\
\hline
$n$) & 0 &  $\frac{3.75-0.5n_f}{\epsilon}$ & $\frac{1.75}{\epsilon}+7.953$ & $\frac{1.75}{\epsilon}+(7.96\pm0.02)$
\\
&&$+15.327-1.844 n_f$&&
\\
\hline
C.~T. ($\delta\alpha_s$)
& $\frac{2 {\rm n_f}-33}{4\epsilon}$ & --- & $\frac{2 n_f-33}{4\epsilon}$ &$\frac{2 n_f-33}{4\epsilon}$
\\
\hline
Total
& $\frac{4.5}{\epsilon^2}-\frac{15.05-0.5 n_f}{\epsilon}$ & $\frac{4.5}{\epsilon^2}+\frac{15.05-0.5 n_f}{\epsilon}$ & $11.14(2)-1.844 n_f$ &
\\
&$-21.862(9)$ & $33.003(7)-1.844 n_f$& $3.77(2)$($n_f=4$) & $3.79\pm 0.53$
\end{tabular}
\end{ruledtabular}
\end{table}

In Table~\ref{table-as-v0}, we tabulate the numerical value of $\Gamma^{(1,0)}/(\alpha_s\Gamma^{(0,0)}/\pi)$ for each class of cut diagrams.
Following \cite{Mackenzie:1981sf}, we also include in the class $c$)
the diagrams with insertion of heavy quark mass counterterm.
Moreover, the entry ``C.~T. ($\delta\alpha_s$)" in Table~\ref{table-as-v0} signifies those tree
diagrams with insertion of the counterterm for the strong coupling constant.
The results from Ref.~\cite{Mackenzie:1981sf} are also juxtaposed with our results, which employed the gluon mass $\lambda$
to regularize IR divergences. To expedite the comparison, we replace $\log \lambda^2/m_Q^2$ in \cite{Mackenzie:1981sf}
with $1/\epsilon_{\rm IR}+\log \mu^2/m_Q^2$.
As can be clearly observed from the Table~\ref{table-as-v0}, we find perfect agreement between our and their results for each class of
cut diagrams. Since we are equipped with the modern IBP method tailored for multi-loop (multi-body phase space) integrals
and the more accurate non-Monte-Carlo integrator,
it is conceivable that a much better numerical accuracy can be achieved with respect to that in \cite{Mackenzie:1981sf}.

Summing up the contributions from each class of cut diagrams, we end up with the IR-finite
$\mathcal{O}(\alpha_sv^0)$ correction to the perturbative decay rate:
\beq
\label{decay-pert-1-0}
\Gamma^{(1,0)}(Q\overline{Q}(^3S_1)\to 3g)=\Gamma^{(0,0)}(Q\overline{Q}(^3S_1)\to 3g)\frac{\alpha_s}{\pi}\bigg[\frac{3}{4}\beta_0\ln\frac{\mu_R^2}{4E^2}
+11.14(2)-1.8436 n_f\bigg],
\eeq
where $\beta_0=11/3C_A-2/3 n_f$ denotes the one-loop coefficient of the QCD $\beta$ function,
with $n_f$ signifying the number of the active light flavors.
In this work, we take $n_f=3$ for $J/\psi$ decay and $n_f=4$ for $\Upsilon$ decay.
$\mu_R$ in \eqref{decay-pert-1-0} represents the renormalization scale, whose explicit occurrence
is demanded by the renormalization group invariance.

Substituting \eqref{decay-pert-1-0} and \eqref{NRQCD:pert:Matrix:elements:v0} into \eqref{perturbative:matching:formula},
setting $E=m_Q$, we then deduce the SDC at ${\cal O}(\alpha_s^0 v^{0})$:
 \beq
  \label{sdcs-1-0}
F_1^{(1)}({}^3S_1) =F_1^{(0)}({}^3S_1)\frac{\alpha_s}{\pi}\left[\frac{3}{4}\beta_0\ln\frac{\mu_R^2}{4m_Q^2}
+11.14(2)-1.8436 n_f\right],
\eeq
which is compatible with the result in literature~\cite{Mackenzie:1981sf,Bodwin:1994jh,Bodwin:2013zu},
yet bears a much smaller error.

\subsection{SDC at $\mathcal{O}(\alpha_sv^2)$}

We proceed to determine the numerical value of $G_1^{(1)}({}^3S_1)$, by far the unknown SDC at $\mathcal{O}(\alpha_sv^2)$.

\begin{table}
\caption{$\mathcal{O}(\alpha_sv^2)$ contribution to the $\Gamma^{(1,1)}/(\alpha_s v^2 \Gamma^{(0,0)}/\pi)$ from each class of cut diagrams.
We have set the scale affiliated with dimensional regularization to be $\mu=2E$.}.
\begin{ruledtabular}
\label{table-as-v2}
\begin{tabular}{c|c|c|c}
Cut diagrams  & Virtual corr. & Real corr. & Virtual+Real corrs.
\\
\hline
$a$) & $\frac{8.641}{\epsilon}+{33.777}$ & --- & $\frac{8.641}{\epsilon}+{33.777}$
\\
\hline
$b$) & $\frac{0.720}{\epsilon
   }+{1.797}$ & --- & $\frac{0.720}{\epsilon}+{1.797}$
\\
\hline
$c$) & $\frac{5.761}{\epsilon
   }+{5.992}$ & --- & $\frac{5.761}{\epsilon
   }+{5.992}$
\\
\hline
$d$) & $\frac{0.360}{\epsilon}+{0.699}$ & --- & $\frac{0.360}{\epsilon}+{0.699}$
\\
\hline
$e$) & $-1.570$ & --- & $-1.570$
\\
\hline
$f$) & $-\frac{3.983}{\epsilon}+{26.390}$ & --- & $-\frac{3.983}{\epsilon}+{26.390}$
\\
\hline
$g)$-$j$) & $-\frac{3.700}{\epsilon^2}-\frac{27.095}{\epsilon}+{42.489}$ & $\frac{3.700}{\epsilon^2
   }-\frac{2.070}{\epsilon}-{125.918}$ & $-\frac{29.164}{\epsilon}-{83.429}$
\\
\hline
$k$) & $\frac{23.143}{\epsilon^2}+\frac{77.197}{\epsilon}+{123.118}$ & $-\frac{23.143}{\epsilon^2
   }-\frac{77.196}{\epsilon
   }-{140.976}$ & $-{17.857}$
\\
\hline
$l$) & $11.793$ & --- & $11.793$
\\
\hline
$m$) & --- & $4.272$ & $4.272$
\\
\hline
$n$) &--- & $\frac{2.160 n_f-16.202}{\epsilon}$ & $\frac{2.160 n_f-16.202}{\epsilon}$
\\
&&$+9.392 n_f-76.924$ & $+9.392 n_f-76.924$
\\
\hline
C.~T.
& $\frac{-2.160 n_f+35.645}{\epsilon}$ & --- & $\frac{-2.160 n_f+35.645}{\epsilon}$
\\
($\delta \alpha_s$)& $-3.069 n_f+50.636$ & --- & $-3.069 n_f+50.636$
\\
\hline
 Total & $\frac{19.443}{\epsilon^2}+\frac{-2.160 n_f+97.247}{\epsilon}$
 & $-\frac{19.443}{\epsilon^2}+\frac{2.160 n_f-95.469}{\epsilon}$ &  $\frac{16}{9\epsilon}$

\\
&$-3.069 n_f+295.121(3)$ & $+9.3923 n_f-339.545(6)$ & $+6.3235 n_f-44.424(6)$
\end{tabular}
\end{ruledtabular}
\end{table}

In Table~\ref{table-as-v2}, we tabulate the individual contribution from each class of cut diagrams to the perturbative
hadronic decay rate. The numbering convention for the Feynman diagrams is the same as Table~\ref{table-as-v0}.
Upon summing all the individual contribution in Table~\ref{table-as-v2}, and
implementing standard renormalization procedure,
we obtain the following inclusive decay rate in QCD side
at $\mathcal{O}(\alpha_sv^2)$:
\bqa
\label{decay-pert-1-1}
\Gamma^{(1,1)}\left(Q\bar{Q}(^3S_1)\to 3g\right) &=&\Gamma^{(0,0)}(Q\bar{Q}(^3S_1)\to 3g)\frac{\alpha_sv^2}{\pi}\bigg[\frac{24-7\pi^2}{12(\pi^2-9)}\frac{3}{4}\beta_0\ln\frac{\mu_R^2}{4E^2}
\nn
\\
&+&
\frac{16}{9}\left({1\over \epsilon_{\rm IR}}+\ln\frac{\mu_\Lambda^2}{4E^2}\right)-44.424(6)+6.3235 n_f\bigg].
\eqa
Clearly, the occurrence of $G_1^{(0)}({}^3S_1) \beta_0\ln\mu_R$ reflects the $\mu_R$-independence of the
decay rate, which is demanded by the standard renormalization group equation.

A salient trait in \eqref{decay-pert-1-1} is the occurrence of an uncancelled single IR pole.
We assign the symbol $\mu_\Lambda$ to denote the scale accompanied with this single IR pole, which
has very different origin from the QCD renormalization scale $\mu_R$.

The uncanceled single IR pole in \eqref{decay-pert-1-1}, which arises at
the hard region at ${\cal O}(\alpha_s v^2)$, is actually not surprising at all.
In fact, this symptom was first discovered in the ${\cal O}(\alpha_s v^2)$ corrections for
$J/\psi\to e^+e^-$~\cite{Luke:1997ys} (see also \cite{Bodwin:2008vp}),
$\eta_c\to \gamma\gamma$~\cite{Jia:2011ah},  $J/\psi\to 3\gamma$~\cite{Feng:2012by}, and $\eta_c$ total hadronic width~\cite{Guo:2011tz}.
Physically, this unremoved IR pole indicates the breakdown of
the color transparency once beyond the LO in $v$. This IR divergence can be factored in the
renormalized NRQCD matrix element,
so that the ${\cal O}(\alpha_s v^2)$ SDC will explicitly depend on the NRQCD factorization scale $\mu_\Lambda$,
which naturally ranges from $m_Q v$ to $m_Q$.

Expanding $E$ in (\ref{decay-pert-1-1}) around $m_Q$ in power series of ${\bm q}^2/m_Q^2$, then comparing (\ref{decay-pert-1-1}) with (\ref{perturbative:matching:formula}), we finally obtain the desired SDC in $\overline{\rm MS}$ factorization scheme:
\bqa
\label{sdcs-1-1}
G_1^{(1)}({}^3S_1)&=&F_1^{(0)}({}^3S_1)\frac{\alpha_s}{\pi}\bigg[\frac{132-19\pi^2}{12(\pi^2-9)}\frac{3}{4}\beta_0\ln\frac{\mu_R^2}{4m_Q^2}
+\frac{16}{9}\ln\frac{\mu_\Lambda^2}{4m_Q^2}\nn\\
&-&63.82(2)+8.6671 n_f\bigg].
\eqa

Equation \eqref{sdcs-1-1} constitutes the major new finding of this work.
It is interesting to note that the SDC at ${\cal O}(\alpha_s v^2)$
contains the same $\ln \mu_\Lambda$ term as the case for $J/\psi\to 3\gamma$~\cite{Feng:2012by}.
However, \eqref{sdcs-1-1} also contains a negative non-logarithmic constant, in sharp
contrast with what is found for $J/\psi\to 3\gamma$~\cite{Feng:2012by}.

\section{Phenomenological analysis~\label{Section:Phenomenological:Analysis}}

Having all the SDCs through ${\cal O}(\alpha_s)$ at hand,
we are ready to conduct a detailed phenomenological analysis
based on the NRQCD factorization formula \eqref{NRQCD:factorization:formula}.

\subsection{Phenomenology for $J/\psi(\Upsilon)\to {\rm LH}$~\label{SubSection:Phenomenology:jpsi:to:LH}}

Before launching into concrete numerics, we would like to first develop some intuition
on the relative importance of the various corrections.
Setting $\mu_R=\mu_\Lambda=m_Q$, we then obtain
\begin{subequations}
\begin{align}
& \Gamma[J/\psi\to {\rm LH}]=
0.0716\frac{\alpha_s^3\langle \mathcal{O}_1({}^3S_1)\rangle_{J/\psi}}{m_c^2}\bigg[
1-5.32\langle v^2\rangle_{J/\psi}-1.19\alpha_s+3.03\alpha_s\langle v^2\rangle_{J/\psi}
\bigg],
\label{Jpsi:hadronic:decay:rate:intuition}
\\
& \Gamma[\Upsilon\to {\rm LH}]=
0.0716\frac{\alpha_s^3\langle \mathcal{O}_1({}^3S_1)\rangle_{\Upsilon}}{m_b^2}\bigg[
1-5.32\langle v^2\rangle_{\Upsilon}-1.56\alpha_s+4.61\alpha_s\langle v^2\rangle_{\Upsilon}
\bigg].
\end{align}
\label{decay-rate-num-1}
\end{subequations}
To condense the notation, we have introduced the dimensionless ratio $\langle v^2\rangle_{V}$ to characterize
the relativistic correction:
\bqa
\langle v^2\rangle_{V}\equiv \frac{\langle 0| \chi^\dagger{\bm
\sigma}\cdot{\bm\epsilon}^\ast
\big(-\tfrac{i}{2}\overleftrightarrow{\bf{D}}\big)^2
\psi|V\rangle}{m_Q^2\langle 0| \chi^\dagger\bm{\sigma}\cdot
\bm{\epsilon}^\ast\psi|V\rangle}.
\label{v2:definition}
\eqa

To make concrete predictions, we need further specify various input parameters: the value of heavy quark masses,
various NRQCD matrix elements, as well as the running strong coupling constant evaluated around the quarkonium mass.
We choose the quark pole mass to be $m_c=1.4$ GeV and $m_b=4.6$ GeV~\cite{Feng:2012by}.
We take the same value of NRQCD matrix elements for $J/\psi$ as in Ref.~\cite{Bodwin:2007fz}:
\beq
\label{LDME-charm}
\langle \mathcal{O}_1({}^3S_1)\rangle_{J/\psi}=0.446\;{\rm GeV}^3,\qquad\quad \langle v^2\rangle_{J/\psi}=0.223.
\eeq
We also take the values of the NRQCD matrix elements for $\Upsilon(nS)$ from Ref.~\cite{Chung:2010vz}~\footnote{Note that
$\langle v^2\rangle_{\Upsilon(1S)}$ has also been fitted to be $-0.078$ through semi-inclusive decay $\Upsilon(1S)\to D^{*+}X$~\cite{Chen:2011ph},
with the magnitude significantly greater than that given in \cite{Chung:2010vz}.}:
\begin{subequations}
\bqa
\langle \mathcal{O}_1({}^3S_1)\rangle_{\Upsilon(1S)}&=&3.069\;{\rm GeV}^3,\qquad\quad
\langle v^2\rangle_{\Upsilon(1S)}=-0.009\pm 0.003,
\label{Upsilon:1S:LDMEs}
\\
\langle \mathcal{O}_1({}^3S_1)\rangle_{\Upsilon(2S)}&=&1.623\;{\rm GeV}^3,\qquad\quad
\langle v^2\rangle_{\Upsilon(2S)}= 0.09.
\label{Upsilon:2S:LDMEs}
\eqa
\label{Upsilon:1S:2S:NRQCD:LDMEs}
\end{subequations}

We use the package {\tt RunDec}~\cite{Chetyrkin:2000yt} to compute the QCD running coupling constant
to two-loop accuracy:
\beq
\label{as-value}
\alpha_s(m_c)=0.36,\qquad
\alpha_s(m_b)=0.22.
\eeq

\begin{table}
\caption{NRQCD predictions to $J/\psi(\Upsilon(nS)\to {\rm LH}$ at various level of accuracy in $\alpha_s$ and $v$,
by setting $\mu_R=\mu_\Lambda=m_Q$. The last column lists the measured values
taken from {\tt PDG 2020}~\cite{Zyla:2020zbs}.}
\begin{ruledtabular}
\label{table-Numerics:Predictions:V:to:LH}
\begin{tabular}{c|c|c|c|c|c|c|}
\diagbox{V}{$\Gamma$(keV)}{Order} & LO & $\mathcal{O}(v^2)$ & $\mathcal{O}(\alpha_s)$ & $\mathcal{O}(\alpha_sv^2)$ & Total &
Exp.~\cite{Zyla:2020zbs} \\
\hline
$J/\psi$ & $767$ & $-910$ & $-330$ & $187$ & $-286$& $59.5\pm2.7$\\
\hline
$\Upsilon(1S)$ & $108$ & $5$ & $-37$ & $-1$ & $75$&  $44.1\pm1.4$\\
\hline
$\Upsilon(2S)$ & $57$ & $-27$ & $-19$ & $5$ & $15$& $18.8\pm2.0$\\
\end{tabular}
\end{ruledtabular}
\end{table}

Our predictions for the hadronic widths of $J/\psi$ and $\Upsilon(1S,2S)$ are tabulated
in Table~\ref{table-Numerics:Predictions:V:to:LH}.
One readily observes that the $\mathcal{O}(\alpha_sv^2)$ corrections are moderate in magnitude
with respect to $\mathcal{O}(\alpha_s^0 v^2)$ and $\mathcal{O}(\alpha_s v^0)$ corrections,
which may indicate that NRQCD expansion exhibits reasonable convergence behavior both in $\alpha_s$ and $v$
in these channels.
It is interesting to remark that, this pattern is in sharp contrast with the situation for
$J/\psi\to 3\gamma$, where one encounters a very significant positive $\mathcal{O}(\alpha_sv^2)$ correction!

\begin{figure}[htbp]
\centering
\includegraphics[width=0.8\textwidth]{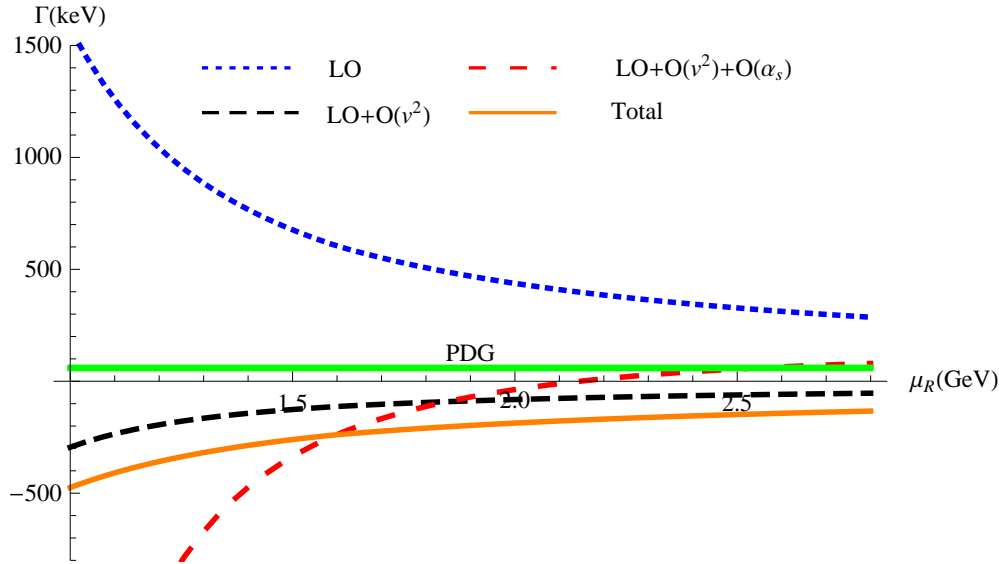}
\caption{Predicted hadronic width of $J/\psi$ (in unit of keV) as a function of
the renormalization scale $\mu_R$ by holding the NRQCD factorization scale fixed, $\mu_\Lambda=m_c$.
Different curves designate the various predictions at various levels of accuracy in $\alpha_s$
and $v$ expansion. \label{Fig:Jpsi:width:depending:on:muR}}
\end{figure}

\begin{figure}[htbp]
\centering
\includegraphics[width=0.49\textwidth]{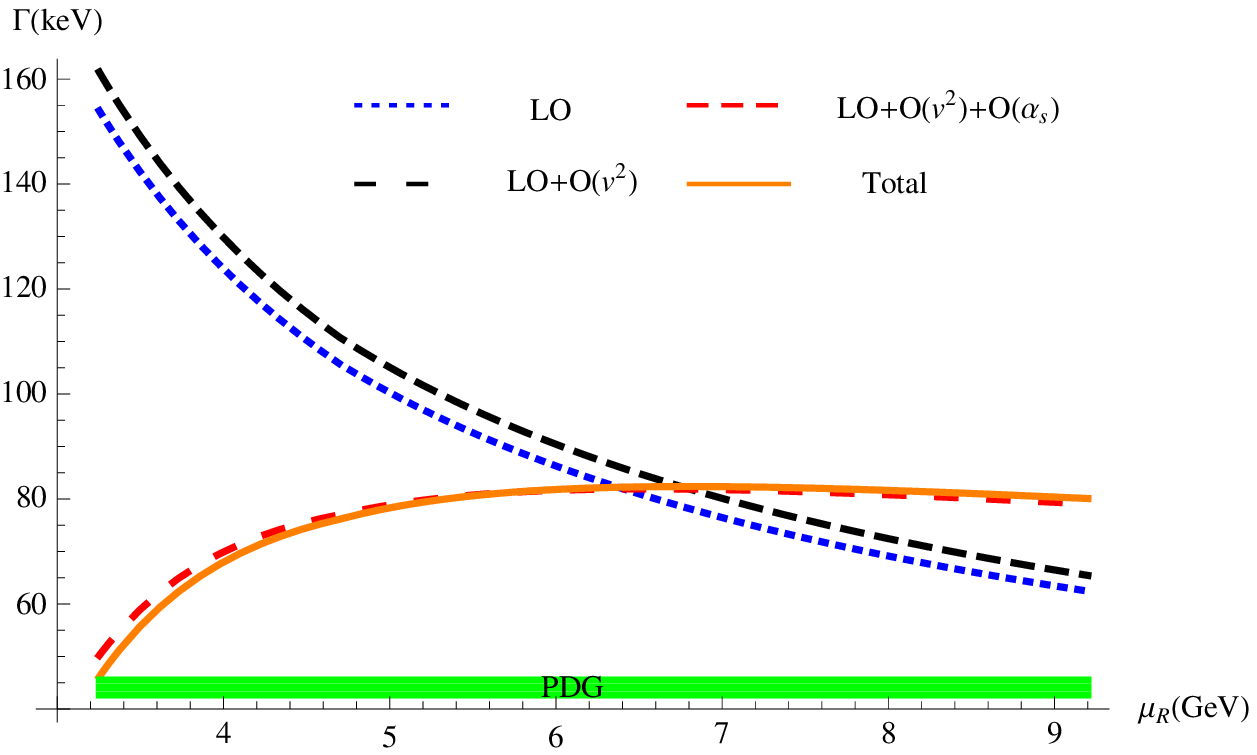}
\includegraphics[width=0.49\textwidth]{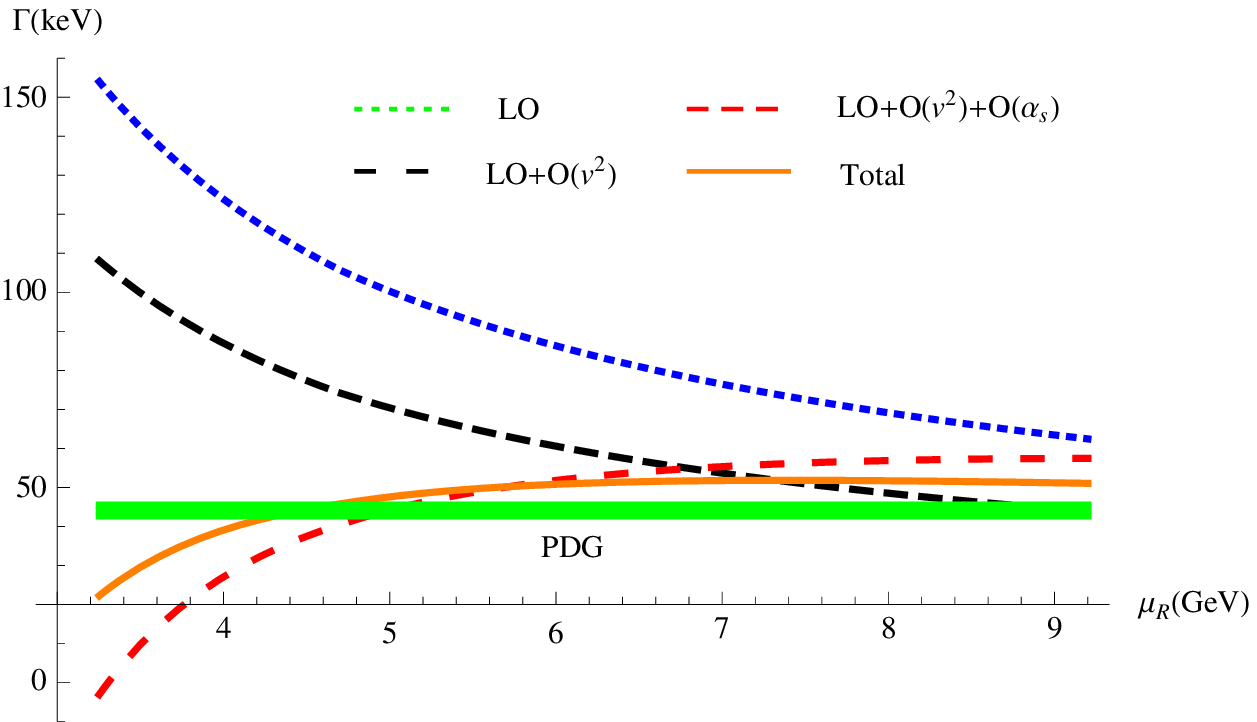}
\caption{Predicted decay rate for $\Upsilon(1S) \to {\rm LH}$ as a function of $\mu_R$
at various levels of accuracy in NRQCD expansion, with $\mu_\Lambda=m_b$.
For the plot on the left panel, we take the value of $\langle v^2\rangle_{\Upsilon(1S)}$ same as in \eqref{Upsilon:1S:LDMEs},
while on the right panel we take $\langle v^2\rangle_{\Upsilon(1S)}=0.056$ in
\eqref{v2:G-K:relation:Upsilon}, which is determined through G-K relation.
\label{fig-upsilon}}
\end{figure}

\begin{figure}[htbp]
\centering
\includegraphics[width=0.8\textwidth]{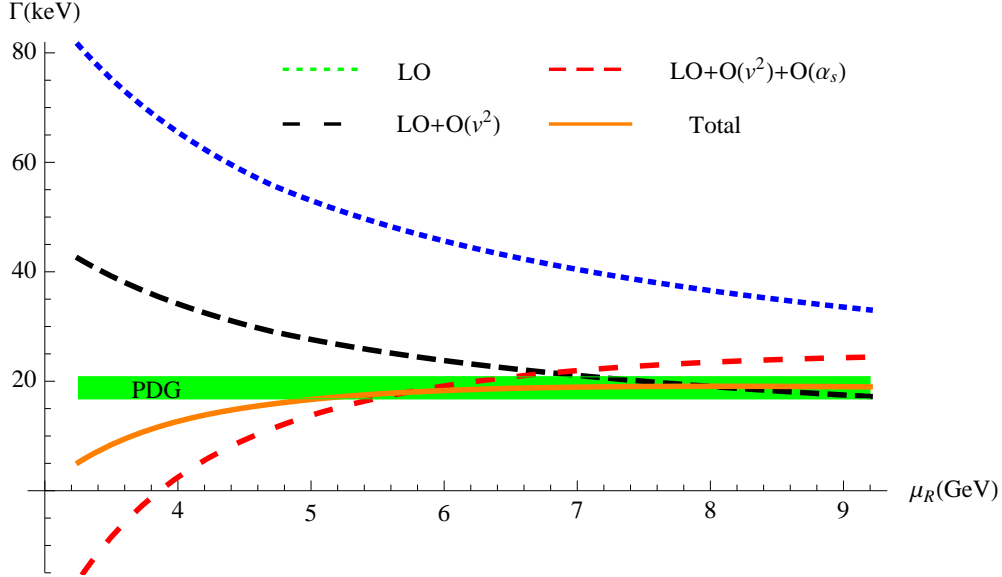}
\caption{Predicted decay rate for $\Upsilon(2S) \to {\rm LH}$ as a function of $\mu_R$
at various levels of accuracy in NRQCD expansion, with $\mu_\Lambda=m_b$.
\label{fig-upsilon2s}}
\end{figure}

As one easily recognizes from Table~\ref{table-Numerics:Predictions:V:to:LH},
a salient feature for $J/\psi \to {\rm LH}$ is the disquietingly negative $\mathcal{O}(\alpha_s^0v^2)$ correction,
which is even greater than LO prediction in magnitude. Even worse, the $\mathcal{O}(\alpha_s v^0)$ radiative correction
further decreases the predicted hadronic width.  Although the new $\mathcal{O}(\alpha_s v^2)$ correction
tends to increase the prediction, unfortunately its effect is not significant enough to bring the predicted hadronic width
to a positive value, so that we are unable to make a meaningful prediction to confront the measurement.

To closely examine this apparent discrepancy between theory and experiment,
we further display in Fig.~\ref{Fig:Jpsi:width:depending:on:muR}
the dependence of $J/\psi$ hadronic decay rate on the renormalization scale $\mu_R$
at various level of accuracy in $\alpha_s$ and $v$ expansion,
with the NRQCD factorization scale $\mu_\Lambda=m_c$ held fixed.
In the reasonable range of $1\;{\rm GeV}<\mu_R<2m_c$, we again observe that
the total decay rate is always {\it negative}.
As mentioned before, the root of this dilemma can be readily traced from \eqref{Jpsi:hadronic:decay:rate:intuition}, that is,
mainly due to the substantially negative $\mathcal{O}(\alpha_s^0 v^2)$ and $\mathcal{O}(\alpha_s v^0)$ corrections and moderately positive $\mathcal{O}(\alpha_s v^2)$ correction.
Thus, we conclude that the state-of-the-art NRQCD prediction ceases to yield a physically meaningful result
for $J/\psi \to {\rm LH}$. How to account for the experimental data from the perspective of NRQCD factorization remains an
open challenge. As a potentially appealing solution, one may conjecture that
the uncalculated ${\cal O}(\alpha_s^2 v^0)$ correction might be significantly positive, which would bring the NRQCD prediction
closer to measured value.
Unfortunately, the bottleneck of the current multi-loop(leg) calculational capability
impedes one to tackle this type of NNLO perturbative correction in the foreseeable future.

From Table~\ref{table-Numerics:Predictions:V:to:LH}, one observes that the ${\cal O}(\alpha_s v^2)$ correction has
rather minor effect for $\Upsilon(1S,2S)\to {\rm LH}$, much less significant than
the ${\cal O}(\alpha_s v^0)$ and ${\cal O}(\alpha_s^0 v^2)$ corrections.
In Fig.~\ref{fig-upsilon} and Fig.~\ref{fig-upsilon2s}, the hadronic decay rates of $\Upsilon(1S,2S)$
are displayed as a function of the renormalization scale $\mu_R$.
As can be easily visualized, though the LO NRQCD prediction exhibits strong $\mu_R$ dependence, after including the corrections
through ${\cal O}(\alpha_s v^2)$, the decay rate becomes much less sensitive to the variation of $\mu_R$, especially when
$\mu_R>m_b$.

The relativistic corrections in $\Upsilon(1S)$ hadronic decay appear to be negligible,
which can be attributed to the highly suppressed relativistic NRQCD matrix elements, as clearly seen
in \eqref{Upsilon:1S:LDMEs}.
We choose two different values for the relativistic NRQCD matrix element
to make predictions for $\Upsilon(1S)\to {\rm LH}$.
When choosing the values specified in \eqref{Upsilon:1S:LDMEs}, the finest NRQCD
prediction for $\Upsilon(1S)$ hadronic width is considerably larger than the measured value,
which can be readily seen in Table~\ref{table-Numerics:Predictions:V:to:LH} and the left panel of Fig.~\ref{fig-upsilon}.

Alternatively, one can estimate the $\langle v^2\rangle_{\Upsilon(1S)}$ according to the
Gremm-Kapustin (G-K) relation~\cite{Gremm:1997dq}:
\beq
\langle v^2\rangle_{\Upsilon(1S)} \approx {M_{\Upsilon(1S)}-2m_b\over m_b}= 0.056,
\label{v2:G-K:relation:Upsilon}
\eeq
once $m_b=4.6$ GeV is chosen.
With such choice, the $\mathcal{O}(\alpha_sv^2)$ correction may reach $10\%$ of the LO decay rate.
Consequently, for $m_b\le \mu_R \le 2m_b$,  the hadronic width of $\Upsilon(1S)$ is predicted to be
in the range $44-51\;{\rm keV}$.
Interestingly,
as one can tell from the right panel of Fig.~\ref{fig-upsilon},
the finest NRQCD prediction is in satisfactory agreement with the measured value: $44.1\pm 1.4\; {\rm keV}$~\cite{Zyla:2020zbs}.

For $\Upsilon(2S)$ hadronic width, both the $\mathcal{O}(\alpha_s v^0)$ and $\mathcal{O}(\alpha_s^0 v^2)$ corrections are
negative and sizeable.
The $\mathcal{O}(\alpha_sv^2)$ correction turns out to be positive, moderate in magnitude.
Incorporating the new piece of $\mathcal{O}(\alpha_sv^2)$ correction, the finest NRQCD prediction for the hadronic width
is in the range $15-19\;{\rm keV}$ for $m_b\le \mu_R \le 2m_b$,  compatible with the measured value
$18.8\pm2.0\; {\rm keV}$~\cite{Zyla:2020zbs}.

\subsection{Phenomenology for $J/\psi(\Upsilon)\to 3\gamma$~\label{SubSection:Phenomenology:jpsi:to:3photon}}

\begin{figure}[htbp]
\centering
\includegraphics[width=0.49\textwidth]{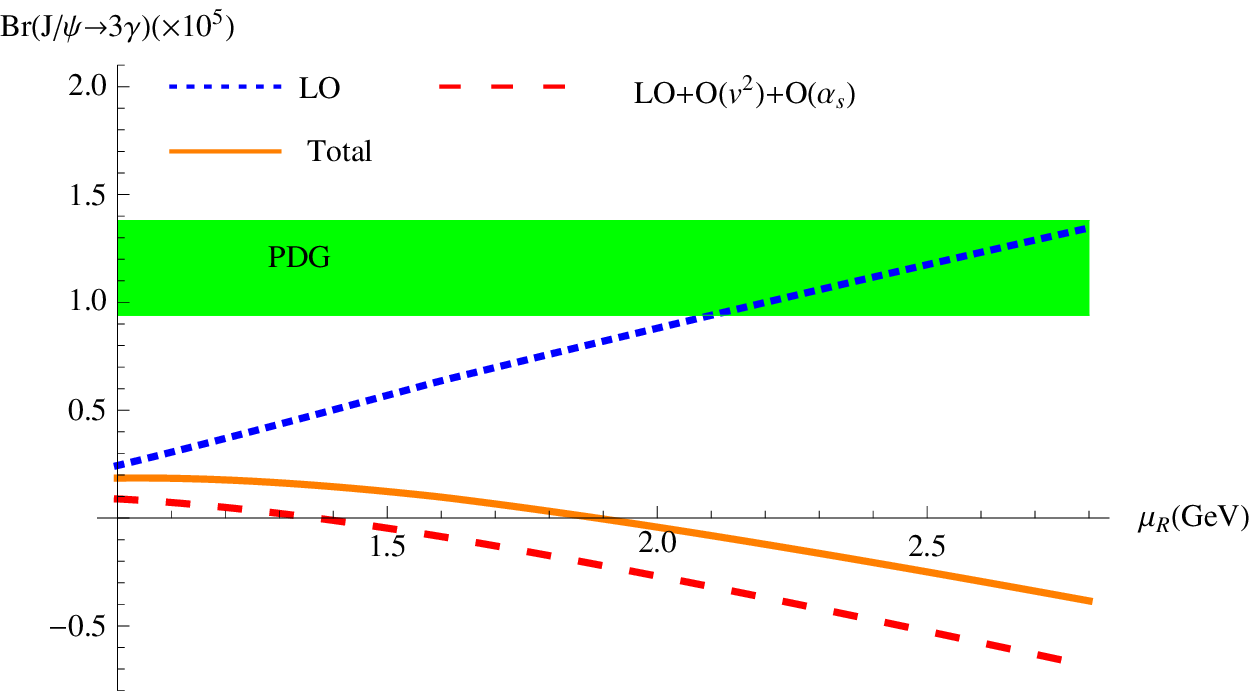}
\includegraphics[width=0.49\textwidth]{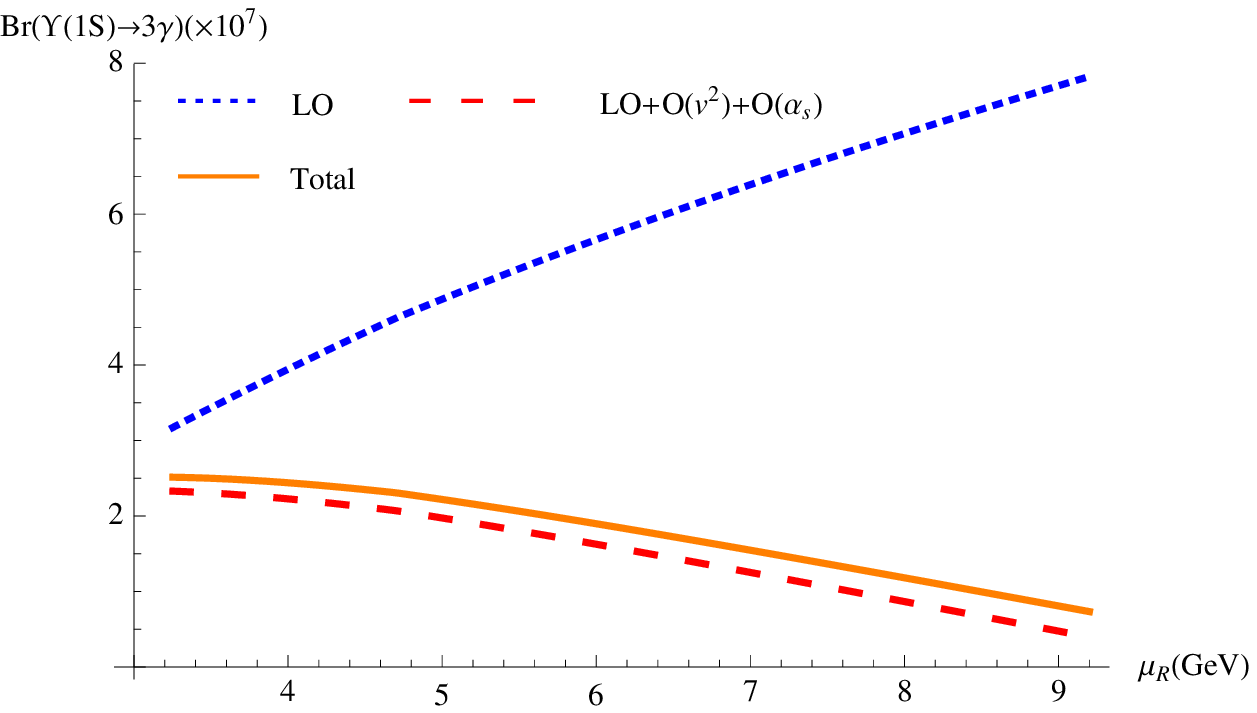}
\caption{NRQCD predictions for the branching fractions of $J/\psi \to 3\gamma$ (left panel) and
$\Upsilon(1S) \to 3\gamma$ (right panel) with various levels of accuracy in $\alpha_s$ and $v$ expansion,
with $\mu_\Lambda$ fixed to be heavy quark mass.
The rare electromagnetic decay $\Upsilon\to 3\gamma$ has not yet been observed experimentally.
We take $\langle v^2\rangle_{\Upsilon(1S)}=0.056$ in \eqref{v2:G-K:relation:Upsilon},
which is inferred from G-K relation.
\label{Fig:Ratio:V:EM:Decay:3photon}}
\end{figure}

The knowledge of the ${\cal O}(\alpha_s v^2)$ correction enables us not only to present
the most complete NRQCD predictions for hadronic widths of vector quarkonia,  but present
the finest NRQCD predictions for the rare electromagnetic decay $J/\psi(\Upsilon)\to 3\gamma$.

The ${\cal O}(\alpha_s v^2)$ corrections to the partial width of $J/\psi(\Upsilon)\to 3\gamma$ were
addressed by the authors some time ago~\cite{Feng:2012by}:
\bqa
\label{decay-rate-jpsi-3-photon}
\Gamma(V\to 3\gamma)&=&\frac{8(\pi^2-9)e_Q^6\alpha^3}{9m_Q^2}\langle \mathcal{O}_1({}^3S_1)\rangle_{V}\bigg\{1-12.630\frac{\alpha_s}{\pi}
\\
&+&\bigg[\frac{132-19\pi^2}{12(\pi^2-9)}+ \bigg(\frac{16}{9}\ln\frac{\mu_\Lambda^2}{m_Q^2}+68.913\bigg)\frac{\alpha_s}{\pi}\bigg]\langle v^2\rangle_{V}\bigg\},\nn
\eqa
where $e_Q$ represents the electric charge of the heavy quark, and $\alpha$ denotes the fine structure constant.
Curiously, the $\ln \mu_\Lambda$ term in the ${\cal O}(\alpha_s v^2)$ SDC is identical to that in $G_1^{(1)}({}^3S_1)$ in \eqref{sdcs-1-1}.

Dividing \eqref{decay-rate-jpsi-3-photon} by \eqref{NRQCD:factorization:formula}, and plugging the result for various SDCs
through ${\cal O}(\alpha_s v^2)$ that are tabulated in Section~\ref{SDC:numerical},
we then obtain the state-of-the-art NRQCD predictions for the branching fractions of $J/\psi(\Upsilon)\to 3\gamma$.
In the spirit of NRQCD expansion, we further expand the ratio in power series of $\alpha_s$ and $v^2$ and obtain
\begin{subequations}
\label{Branching:fractions:V:to:3:gamma}
\bqa
\label{br-jpsi-3-photon}
{\rm Br}(J/\psi\to 3\gamma)&=& {\rm Br}(J/\psi\to {\rm LH}) \times {\Gamma(J/\psi\to 3\gamma)\over \Gamma(J/\psi\to {\rm LH})}
\\
&=& 0.641 \frac{\alpha^3}{\alpha_s^3}\bigg[0.948-\frac{\alpha_s}{\pi}\bigg(
8.42+6.40\ln\frac{\mu_R^2}{m_c^2}
\bigg)+\frac{\alpha_s}{\pi}\langle v^2\rangle_{J/\psi}\;11.51\bigg],
\nn
\\
\label{br-upsilon-3-photon}
{\rm Br}(\Upsilon(1S)\to 3\gamma)&=&{\rm Br}(\Upsilon\to {\rm LH})\times {\Gamma(\Upsilon\to 3\gamma)\over \Gamma(\Upsilon\to {\rm LH})}
\\
&=& 0.817 \frac{\alpha^3}{\alpha_s^3}\bigg[0.0148-\frac{\alpha_s}{\pi}\bigg(
0.115+0.093\ln\frac{\mu_R^2}{m_b^2}
\bigg)+\frac{\alpha_s}{\pi}\langle v^2\rangle_{\Upsilon(1S)}\;0.197\bigg].
\nn
\eqa
\end{subequations}
To reduce the theoretical error, we take ${\rm Br}(J/\psi\to {\rm LH})=64.1\%$ and ${\rm Br}(\Upsilon(1S)\to {\rm LH})=81.7\%$ as
experimental input~\cite{Zyla:2020zbs}.

As is evident in \eqref{Branching:fractions:V:to:3:gamma}, The LO NRQCD matrix element has canceled in the ratio.
More interestingly, the ${\cal O}(\alpha_s^0 v^2)$ correction has also completely disappear in the ratio.
Furthermore, the explicit logarithmic dependence on NRQCD factorization scale $\mu_\Lambda$ has also disappeared.

In Fig.~\ref{Fig:Ratio:V:EM:Decay:3photon} we display our predictions for the branching fractions of $J/\psi \to 3\gamma$ (left panel)
and $\Upsilon(1S) \to 3\gamma$ (right panel) at various levels of accuracy in $\alpha_s$ and $v$ expansion,
with $\mu_\Lambda=m_Q$. The sensitivity of the branching fractions to $\mu_R$ seems not to improve much after incorporating
the higher order corrections.
This may be attributed to the sizeable radiative corrections and the $\alpha_s^3$ factor in the denominator.
Setting $\mu_R=m_c$, the finest NRQCD prediction is ${\rm Br}(J/\psi\to 3\gamma)=1.45\times10^{-6}$,
about one order of magnitude smaller than the measured value $1.16\times 10^{-5}$.
As can be seen in Fig.~\ref{Fig:Ratio:V:EM:Decay:3photon},
varying the renormalization scale in the range $1\; {\rm GeV}\le \mu_R \le 3\;{\rm GeV}$,
we always observe that ${\rm Br}(J/\psi\to 3\gamma)$ is far smaller than the experimental measurement.
The resolution to this dilemma may need call for including
further higher order relativistic corrections~\cite{Bodwin:2007fz}, or incorporating the ${\cal O}(\alpha_s^2 v^0)$
correction.
Such a study appears to be extremely challenging on technical ground, which is beyond the scope of this work.

In a similar fashion, with the aid of \eqref{br-upsilon-3-photon},
we can predict ${\rm Br}(\Upsilon(1S)\to 3\gamma)=2.06^{+0.20}_{-1.55}\times10^{-7}$ and ${\rm Br}(\Upsilon(2S)\to 3\gamma)=1.78^{+0.11}_{-0.99}\times10^{-7}$, where the uncertainties are estimated by
varying $\mu_R$ from $m_b/2$ to $2m_b$. The uncertainties looks significant due to strong $\mu_R$ dependence.
Alternatively, if we choose $\langle v^2\rangle_{\Upsilon(1S)}=0.056$ as indicated by the G-K relation,
the branching fraction shifts to ${\rm Br}(\Upsilon(1S)\to 3\gamma)=2.33^{+0.18}_{-1.44}\times10^{-7}$.
It is a clear sign that the relativistic effect is less important for bottomonium case, and the
branching fraction appears to be insensitive to the relativistic corrections. We hope future experiments for
$\Upsilon$ rare electromagnetic decay can test our predictions.

\section{Summary\label{summary}}

In summary, we have computed the ${\mathcal{O}(\alpha_sv^2)}$ corrections to $J/\psi(\Upsilon)$ hadronic decays,
precisely deducing the corresponding SDCs. Unlike the case in $J/\psi(\Upsilon)\to3\gamma$, the corrections are moderate,
therefore the expansion in $v^2$ exhibits a decent convergence behavior.
We find that the theoretical predictions for the decay width of $J/\psi\to {\rm LH}$ is negative through
$\mathcal{O}(\alpha_sv^2)$, which is certainly unphysical and can be attributed to the sizable and negative relativistic corrections.
On the other hand, we find the theoretical predictions for $\Upsilon(1S, 2S)$ hadronic decays
are consistent with the experimental data, if the NRQCD matrix elements $\langle v^2\rangle_{\Upsilon(nS)}$ are
appropriately chosen. Incorporating the formulas in the Ref.~\cite{Feng:2012by}, we derive the branching fraction for $J/\psi(\Upsilon)\to3\gamma$ accurate up to ${\mathcal{O}(\alpha_sv^2)}$. Although the ${\mathcal{O}(v^2)}$ relativistic corrections
get exactly cancelled, we observe a strong $\mu_R$ dependence in the branching ratio.  Once again, we find the theoretical prediction for ${\rm Br}(J/\psi\to 3\gamma)$ can not explain the experimental measurement.
In our opinion, the theoretical incompetence to precisely predicting $J/\psi$ decay in NRQCD factorization approach
may be attributed to the not-yet-known higher-order relativistic and perturbative corrections.

\begin{acknowledgements}
We would like to thank A.~V.~Smirnov for helping us to reduce the
tensor integrals encountered in four-particle phase-space integration into a set of MIs before the
\textsf{C++} version of \textsf{FIRE} was officially released.
The work of W.-L. S. is supported by the National Natural Science Foundation
of China under Grants No. 11975187 and the Natural Science Foundation of ChongQing under Grant No. cstc2019jcyj-msxm2667.
The work of F.~F. is supported by the National Natural
Science Foundation of China under Grant No. 11875318,
No. 11505285, and by the Yue Qi Young Scholar Project
in CUMTB.
The work of Y.~J. is supported in part by the National Natural Science Foundation of China
under Grants No.~11925506, 11875263,
No.~11621131001 (CRC110 by DFG and NSFC).

\end{acknowledgements}

\appendix

\section{Tactics in dealing with four-body phase space integration~\label{phase:space:integration}}

In this Appendix, we present some technical details on how we parameterize the three-particle and four-particle phase space integrals in $d$ dimensions. Our central goal is to eliminate the Heaviside step function in the four-particle phase space integration through some
trick, so that we can choose some highly precise numerical integrator other than the widely-used, but less accurate,
adaptive Monte Carlo integrator.

We start with the Born-order process for $J/\psi(\Upsilon)\to {\rm LH}$, where the decay product
is comprised of three gluons.
We define the following Lorentz invariants:
 \bqa
\label{lorentz-invariables}
&&P\cdot k_1=2x_1E^2,\;\;\; P\cdot k_2=2x_2E^2,\;\;\; P\cdot k_1=2(2-x_1-x_2)E^2,\nn\\
&&k_1\cdot k_2=2E^2(x_1+x_2-1),\;\;\; k_1\cdot k_3=2E^2(1-x_2),\;\;\; k_2\cdot k_3=2E^2(1-x_1),
\eqa
where the momenta $P$ and $k_i\,(i=1,2,3)$ have been defined in Section~\ref{shortcut:deduce:SDC}.
The dimensionless variables $x_1$ and $x_2$ are constrained to be $1\le x_1+x_2\le 2$ by energy conservations.
It is useful to change variables from $(x_1,x_2)$ to $(x,y)$
so that the integration range of the new variables lie in a square~\cite{Petrelli:1997ge}:
\bqa
x_1=x,\;\;\;  x_2=1-x y,
\eqa
with $0\le x,y\le 1$.

It is then straightforward to parameterize the phase space integral of
three massless particle in $d=4-2\epsilon$ space-time
dimensions as
\bqa
\label{3-body}
 \int \!\!d\Phi_3 &=& \int\!\!\!\frac{d^{d-1}k_1}{2k_1^0(2\pi)^{d-1}}\frac{d^{d-1}k_2}{2k_2^0(2\pi)^{d-1}} \frac{d^{d-1}k_3} {2k_3^0(2\pi)^{d-1}}(2\pi)^d\delta^d(P-k_1-k_2-k_3)
\nn\\
&& =
 \frac{(4E^2)^{1-2\epsilon}}{128\pi^3}\frac{(4\pi)^{2\epsilon}}{\Gamma(2-2\epsilon)}\int_0^1\!
 dx\, dy\: x^{1-2\epsilon}\bigg[y(1-x)(1-y)\bigg]^{-\epsilon}.
\eqa

Calculation of the real corrections demands considering the $J/\psi(\Upsilon)$ decay into four massless partons.
The four-particle phase space integration in dimensional regularization is somewhat intriguing.
We employ the parametrization scheme introduced in Ref.~\cite{GehrmannDe Ridder:2003bm}.
The Mandelstam variables are defined as $s_{ij}=k_i \cdot k_j$, which obey
$s_{12}+s_{13}+s_{14}+s_{23}+s_{24}+s_{34}=4E^2$ constrained by energy conservation.
It is convenient to introduce a set of dimensionless variables $x_i$ through rescaling the
Mandelstam invariants by
\bqa
\label{4-body-variables}
&& x_1=\frac{s_{12}}{4E^2}, \qquad x_2=\frac{s_{13}}{4E^2}, \qquad x_3=\frac{s_{23}}{4E^2},
\\
&& x_4=\frac{s_{14}}{4E^2},\qquad  x_5=\frac{s_{24}}{4E^2}, \qquad x_6=\frac{s_{34}}{4E^2}.
\nn\eqa

In terms of the dimensionless variables in \eqref{4-body-variables},
the massless four-body phase space can be parameterized as~\cite{GehrmannDe Ridder:2003bm}
\bqa
\label{4-body-1}
\int \!\!d\Phi_4 &=& \int\!\!\!\frac{d^{d-1}k_1}{2k_1^0(2\pi)^{d-1}}\frac{d^{d-1} k_2}{2k_2^0(2\pi)^{d-1}} \frac{d^{d-1}k_3}{2k_3^0(2\pi)^{d-1}}\frac{d^{d-1} k_3}{2k_4^0(2\pi)^{d-1}}(2\pi)^d\delta(P-k_1-k_2-k_3-k_4)
\nn\\
&=&  C_\Gamma(4E^2)^{\frac{3d}{2}-4}\int_0^1\!\! \prod_{j=1}^6 dx_j \delta\left(1-\sum_{i=1}^{6}{x_i}\right)[-\lambda(x_1x_6,x_2x_5,x_3x_4)]^{\frac{d-5}{2}}\Theta(-\lambda),
\eqa
where $\lambda(x,y,z)=x^2+y^2+z^2-2(x y+x z+y z)$ is
the K\"allen function, $\Theta$ represents the Heaviside step function,
and the volume factor $C_\Gamma$ is given by
\beq
C_\Gamma=(2\pi)^{4-3d}V(d-1)V(d-2)V(d-3)2^{1-2d},
\eeq
with $V(d)=\frac{2\pi^{d/2}}{\Gamma(d/2)}$ designating the area of the unit sphere imbedded in $d$-dimensional space.

The presence of Heaviside step function in \eqref{4-body-1} generally renders the integration boundary highly irregular,
so that one has to resort to the Monte Carlo recipe for numerical integration,
with some intrinsic limitation on integration accuracy.
In fact, we can manage to eliminate the Heaviside function through a clever trick,
so that we can employ more accurate numerical integrator other than the Monte Carlo algorithm.

First let us explicitly write down the K\"allen function in \eqref{4-body-1},
abbreviated by $\lambda$ for simplicity:
\bqa
\lambda\equiv\lambda(x_1 x_6,x_2 x_5,x_3 x_4) &=& x_3^2 x_4^2+x_2^2 x_5^2+x_1^2 x_6^2-2(x_2x_3x_4x_5+x_1x_2x_5x_6+x_1x_3x_4x_6).
\nn
\\
\eqa

With the aid of Cheng-Wu theorem~\cite{Cheng-Wu-1,Bjoerkevoll:1992cu,Smirnov_Book},
we have the freedom to pull the variables $x_1$, $x_2$ and $x_3$ outside the $\delta$-function when carrying out phase space integration
\eqref{4-body-1}.
Consequently, the integration range of $x_1$, $x_2$ and $x_3$ then become $[0,\infty)$, essentially unbounded.
We can freely rename the variables:
\bqa
x_1 \to \frac{x_1}{x_6}, \quad x_2 \to \frac{x_2}{x_5}, \quad x_3 \to \frac{x_3}{x_4}.
\eqa
Now the K\"allen function reduces to
\bqa
\lambda &\to& x_1^2 + x_2^2+x_3^2 -2(x_1 x_2 +x_2 x_3 +x_1 x_3) \nonumber\\
&=& (x_1 - x_2)^2 + x_3^2 - 2x_1 x_3 - 2x_2 x_3.
\eqa
We can divide the integration range for $x_1$ and  $x_2$ into two sectors: $x_1\ge x_2$ and $x_1< x_2$.
For the first sector $x_1\ge x_2$, it is convenient to further change the variables:
\beq
 x_2\to \frac{x_2}{2},\qquad x_1\to x_1+\frac{x_2}{2}.
\label{variable:change:2a}
\eeq
For the second sector $x_1< x_2$, we can also make variable change:
\bqa
x_1\to \frac{x_2}{2}\qquad x_2\to x_1+\frac{x_2}{2}.
\eqa
The advantage of changing variables is to ensure that
the support of the new variables are within a square, {\it e.g.},
both integration ranges of new $x_1$ and $x_2$ lie in $[0,\infty)$.

Without loss of generality, we choose the first sector to illustrate our recipe.
After changing variables in line with \eqref{variable:change:2a},
the K\"allen function now simplifies to
\bqa
\lambda=(x_1-x_3)^2-2x_2x_3.
\eqa
Similarly, we can further break the integration into two sectors: $x_1\ge x_3$ and $x_1<x_3$.
For the $x_1\ge x_3$ sector,  we are free to continue to rename the variables:
\beq
x_3\to \frac{x_3}{2}\qquad x_1\to x_1+\frac{x_3}{2}.
\label{variable:change:3a}
\eeq
And for the $x_1< x_3$ sector, we change the variables as
\beq
x_1\to \frac{x_3}{2},\qquad x_3\to x_1+\frac{x_3}{2}.
\eeq
Now the integration range of the new  $x_1$ and $x_3$ variables lie in $[0,\infty)$.

We can repeat the game of dissecting the integral into two sub-sectors.
For concreteness, let us concentrate on the sub-sector $x_1\ge x_3$.
After changing variables as advised in \eqref{variable:change:3a},
the K\"allen function now reduces to
\beq
\lambda=x_1^2-x_2x_3.
\label{Lambda:simplifying}
\eeq
We further rename the variable $x_2$ as
\bqa
x_2\to \frac{x_1x_2}{x_3},
\eqa
so that $\lambda$ in \eqref{Lambda:simplifying} reduces to
\bqa
\lambda=x_1(x_1-x_2).
\eqa

One more again, we can dissect the integration range into two sectors: $x_1\ge x_2$ and $x_1<x_2$.
As indicated by the  Heaviside function $\Theta(-\lambda)$ in \eqref{4-body-1},
the first sub-sector ($x_1\ge x_2$) apparently make vanishing contribution.
Therefore, only the second sub-sector with $x_1<x_2$ survives.
We further change the variables as follows:
\beq
x_1\to \frac{x_2}{2},\qquad x_2\to x_1+\frac{x_2}{2}.
\eeq
So the integration range of the new $x_1$ and $x_2$ variables also
lies in $[0,\infty)$.
Meanwhile, the K\"allen function reduces to $\lambda=-\frac{x_1x_2}{2}$.
Since the Heaviside function $\Theta(-\lambda)$ remains 1 in this prescribed integration support,
it can be safely dropped.
This sequences of manipulation can be recursively applied to any integration sector.

At the last step, we can invoke Cheng-Wu theorem again to put $x_1$, $x_2$ and $x_3$ back inside the $\delta$-function.

Through the aforementioned manipulations, we are able to
eliminate the Heaviside $\Theta$ function from the four-particle phase space integral.
Since the support of the multi-dimensional integration variables is inside a regular hypercube,
we can resort to any efficient numerical integrator, {\it e.g.}, the parallelized integrator {\tt HCubature}~\cite{HCubature}
for high-precision numerical integration.

As a test, we apply sector decomposition method together with our parametrization to numerically
calculate five MIs of the massless four-body phase-space type (labelled by $R_4$, $R_6$, $R_{8,a}$, $R_{8,b}$ and $R_{8,r}$
in \cite{GehrmannDe Ridder:2003bm}).
These five MIs have been analytically worked out in 2003~\cite{GehrmannDe Ridder:2003bm},
with which we find exquisite numerical agreement.



\end{document}